# Supportive psychotherapy on insomnia induced by COVID-19; Evaluation of patients and hospital staff

Atieh Sadeghniiat-Haghighi[1]

Arezu Najafi[2]

Khosro Sadeghniiat Haghighi[3]

Arghavan Shafiee-Aghdam [4]

Farzan Vahedifard[5]

Fatemeh Hoshyar Zare[6]

Maryam Tolouei Shivyari[7]

Mohammad Tolouei[8]

[1] Department of Psychiatry, Roozbeh Hospital, Tehran University of Medical Science

[2] Occupational Sleep Research Center, Tehran University of Medical Sciences

3 Professor. Tehran University of Medical Sciences, Head of Occupational Sleep Research Center

4 Department of Psychiatry, Razi Hospital, University of Social Welfare and Rehabilitation Sciences, Tehran, Iran.

5 Department of Diagnostic Radiology and Nuclear Medicine, Rush Medical College.

6 Neuropsychiatry Research Team , Iran Sleep Disorders Clinic, Tehran, Iran.

7 Psychometery MSc, Islamic Azad University, Tehran, Iran.

8 Neuropsychiatry Research Team , Iran Sleep Disorders Clinic, Tehran, Iran.

# Supportive psychotherapy on insomnia induced by COVID-19; Evaluation of patients and hospital staff

**Abstract:**

**Introduction:** COVID-19, a global pandemic, has increased COVID-19-related stress, anxiety, and sadness among individuals. Insomnia affects a significant proportion of the population, with diagnostic criteria ranging from 6% to 10%. This study aimed to evaluate the efficacy of supportive psychotherapy in managing COVID-19-induced insomnia in patients and hospital staff.

**Method:** The study aimed to evaluate the effectiveness of supportive care provided by a psychiatrist in reducing anxiety and insomnia among COVID-19 patients through a before-and-after design. A total of 18 hospitalized patients with COVID-19 and experiencing symptoms of anxiety or insomnia were included. The treatment consisted of four sessions of individual or group Cognitive-Behavioral Therapy for Insomnia (CBT-I), with each session lasting between 20 minutes and an hour. Baseline assessments of sleep efficiency, severity of insomnia, and symptoms of generalized anxiety disorder were conducted using questionnaires. These evaluations were repeated at the end of the third session to measure the intervention's effectiveness. Descriptive statistics, paired t-tests for mean comparisons, the McNemar test for qualitative comparisons, and regression models to control for confounding factors were utilized during the data analysis process, employing IBM SPSS Statistics.

**Results:** In the patient group, supportive psychotherapy resulted in significant improvements. Anxiety levels decreased from a mean score of 4.49 to 2.65 (p-value = 0.038). Insomnia severity scores improved significantly, with reductions observed in all three insomnia items (Insomnia 1, 2, and 3). Sleep quality was significantly enhanced, as indicated by an increase in the mean score from 6.44 to 8.55 (p-value = 0.005). Similarly, the staff group demonstrated positive outcomes. Anxiety levels decreased significantly from a mean score of 3.88 to 2.6 (p-value = 0.037). Insomnia severity scores improved, with significant reductions in Insomnia 1 and observed reductions in Insomnia 2 and 3. Sleep quality exhibited improvement, although not statistically significant (p-value = 0.08).

**Discussion:** COVID-19-related concerns have heightened stress and disrupted sleep patterns among patients. Supportive psychotherapy, including basic behavioral therapy training, has shown significant efficacy in improving sleep quality and reducing anxiety in a short duration. These findings highlight the positive impact of cognitive-behavioral therapy for insomnia (CBT-I) in managing COVID-19-induced insomnia for both patients and staff.

**Conclusion:** Supportive psychotherapy, particularly CBT-I, is effective in reducing anxiety, improving insomnia symptoms, and enhancing sleep quality in patients and staff. It offers a valuable approach for managing COVID-19-induced insomnia. Further research with larger sample sizes is needed to confirm and generalize these findings. The positive outcomes of CBT-I have important implications for the well-being of individuals affected by the pandemic.

**Introduction**

In December 2019, the Coronavirus disease 2019 (COVID-19) emerged in China, quickly spreading worldwide and causing a global pandemic. COVID-19 has caused fatalities, disturbed millions of lives and trade, and caused sleep and psychological difficulties as it has expanded swiftly.(1) It is discovered that the prevalence of insomnia rose considerably during the COVID-19 outbreak, with 13.6 percent and 12.5 percent of individuals developing new-onset insomnia and worsening insomnia symptoms, respectively, according to the Insomnia Severity Index. Furthermore, COVID-19-related stress was reported by 17.6% of subjects, and the incidence of anxiety and depression grew considerably.(1, 2)

Around a quarter of current people are unhappy with their sleep time or quality, and the diagnostic criteria for insomnia are 6% to 10% (Ohayon & Reynolds, 2009).(3)

Insomnia can be a primary condition or be comorbid with depression (4, 5) and chronic pain in various settings. +3+Primary insomnia (PI) is a frequent and potentially life-threatening condition that impacts functioning, health, and quality of life. The nature and importance of the underlying neural mechanisms and genetic variables affecting PI are currently unknown. However, it has long been recognized that various psychological and behavioral factors contribute to the persistence of this illness. The best way to think of PI is to make a diagnosis. After all, other primary and secondary causes of sleep disturbance have been ruled out. Despite this, PI may typically be detected by a clinical interview, as expensive and time-consuming laboratory testing is rarely required for insomnia diagnosis. (6)

Drug therapy is the most prevalent method of treating insomnia. Benzodiazepine receptor agonists have shown modest effectiveness in several clinical trials (7, 8). Drugs benefit from quick therapeutic improvement if they are widely available and effective. The risk of side effects, dependency, and tolerance with passing time are disadvantages of drug therapy.(9)

Cognitive-behavioral therapy for insomnia (CBT-I) is an alternative or parallel therapeutic option. CBT-I is a non-pharmacological, multi-strategic treatment technique. The purpose of CBT-I is to identify variables that have the potential to cause long-term insomnia, such as sleep dysregulation, sleep-related anxiety, and sleep disorder behaviors that are all driven by sleep. This is accomplished by using sensory control to develop a learned relationship between bed and sleep, restoring sleep homeostatic regulation with sleep restriction, and modifying worried sleep-related thinking with cognitive reconstruction.(9)

However, to our knowledge, Following the COVID-19 epidemic, various psychological concerns, including anxiety and sleeplessness, have emerged among COVID-19 patients, illnesses that have received little attention in recent years.

The aim of the present study we looked at the impact of CBTI on these individuals and the extent of improvement and effectiveness of this treatment among hospital staff and inpatients.(10)

**Method and Materials**

According to the study's entry requirements, participants were hospitalized at Imam Khomeini Hospital with a Covid-19 diagnosis, plus insomnia or anxiety. After gaining the patient's participation, the psychiatrist gave supportive care to the patient in four consecutive virtual or face-to-face sessions for the sample size. The anxiety and insomnia scores of the patients were evaluated at first and at the end of the third session to assess the success of the intervention and compare them to the pre-intervention score.

Inclusion criteria were as fellow: Age group over 18 years old hospitalized due to COVID-19 who are in the process of recovery and discharge; Vigilance and cooperation in accountability; Satisfaction to participate in the study; At least one day has passed since the hospitalization; Having insomnia and anxiety according to the design questionnaires. Also, the exclusion criteria were: Dissatisfaction with participating in the study; Major psychiatric disorders being treated.

Patients had to be treated with either individual or group CBT-I to be included in the trial. Studies on self-help programs were not allowed since the treatment had to be done by an expert. These studies were included when telephone or internet sessions were used with face-to-face CBTI. Enrollment in the trial did not have a minimum or maximum treatment time.

Finally, 18 patients were included in this study. At the beginning of the study, the baseline status of these conditions was assessed in subjects using a questionnaire on sleep quality, insomnia (Insomnia severity index), and General Anxiety Disorder-7)(GAD7). Individuals who met the inclusion criteria were selected. Supportive psychotherapy was performed for each patient daily, and three sessions and each session lasted 20 minutes to half an hour. The patient's concerns were asked in these sessions, and training was given in stress management, muscle relaxation, and sleep health education. After these interventions, we re-evaluated each patient's sleep and stress questionnaire.

Eventually, the data from the surveys were put into the IBM SPSS Statistics version 22. Quantitative factors were described using mean and standard deviation, whereas qualitative variables were described using frequency and percentage. We used the paired T-test to compare means before and after the interventions. To compare qualitative characteristics before and after the intervention, the McNemar test was utilized. By altering the influence of confounding factors, regression models were used to examine the effect of the intervention.

**Result**

Table 1 represents the demographic information of Group 1, which consists of patients. Here are the key details; Table 2 represents the demographic information of Group 2 (staff), which consists of staff members.

By comparing the two tables, we can observe several differences (Group 1 consists of patients, while Group 2 comprises healthcare staff):

1. Group 1 (patients) has a higher average age (58.6) compared to Group 2 (staff), with an average age of 40.72.

2. The gender distribution differs between the two groups. Patients have a more significant proportion of females (12 out of 18), while Group 2 has more females (14 out of 18).

3. The education levels also differ. Patients have a higher proportion of individuals with a high school education, while Group 2 has a higher proportion with a bachelor's degree.

4. The past medical and psychiatric histories differ slightly between the two groups. Patients have a higher incidence of hypertension, diabetes mellitus, ischemic heart disease, kidney disease, surgery, and lung disease. On the other hand, Group 2 has a higher incidence of hyperlipidemia.

These are the main points of comparison between Table 1 and Table 3, highlighting the demographic differences between the two groups of patients and hospital staff.

*Table 1- Demographic Information for group 1 (patients)*

| Category | Sub-Category | Value |
|---|---|---|
| **Gender** | Male | 6 |
| | Female | 12 |
| **Age (mean)** | | 58.6 |
| **BMI** | | 26.17 |
| **Education** | High School | 5 |
| | Diploma | 7 |
| | Associate Degree | 2 |
| | Bachelor's Degree | 4 |
| **Marriage** | Married | 13 |
| | Single (including 2 divorced) | 5 |
| **Job** | Employed | 12 |
| | Housewife | 5 |
| | Retired | 1 |
| **Past medical history** | HTN | 2 |
| | DM | 2 |
| | IHD | 1 |
| | Kidney disease | 1 |
| | Surgery | 1 |
| | Lung disease | 1 |
| **Past psychiatric history** | Anxiety disorders | 7 |
| | Depression | 3 |

*Table 2 - Demographic Information for group 2 (Stuff)*

| Category | Sub-Category | Value |
|---|---|---|
| Gender | Female | 14 |
| | Male | 4 |
| Age (mean) | | 40.72 |
| BMI | | 25.20 |
| Education | Diploma | 2 |
| | Associate Degree | 2 |
| | Bachelor's Degree | 10 |
| | Master's Degree | 4 |
| Marriage | Married | 11 |
| | Single (including 3 divorced) | 7 |
| Job | Healthcare Staff | 18 |

| Category | Sub-Category | Value |
|---|---|---|
| Past medical history | HTN | 1 |
| | DM | 1 |
| | HLP | 2 |
| Past psychiatric history | Anxiety disorders | 5 |

Table 3 presents the anxiety, insomnia, and sleep quality scores before and after receiving cognitive-behavioral therapy for insomnia (CBT-I) in Group 1 (patients).

1. Anxiety total score: The mean anxiety score significantly decreased from 4.49 (pre-CBT-I) to 2.65 (post-CBT-I) with a p-value of 0.038, indicating a statistically significant improvement in anxiety levels after the intervention.

2. Item 1 anxiety: Although the mean score decreased from 1.94 to 1.27 after CBT-I, the p-value of 0.063 suggests a trend towards improvement, but it did not reach statistical significance.

3. Item 2 anxiety: The mean score significantly decreased from 2.55 to 1.38 (p-value = 0.01), indicating a significant reduction in anxiety symptoms related to this specific item after the intervention.

4. Insomnia 1: The mean score significantly decreased from 2.72 to 1.33 (p-value = 0.001), indicating a significant improvement in insomnia severity.

5. Insomnia 2: The mean score decreased from 1.66 to 0.88 after CBT-I, suggesting a reduction in insomnia symptoms. However, the p-value of 0.03 indicates a moderate level of statistical significance.

6. Insomnia 3: The mean score significantly decreased from 2.44 to 1.5 (p-value = 0.04), indicating a statistically significant improvement in insomnia symptoms related to this specific item.

7. Sleep quality: The mean score significantly improved from 6.44 (pre-CBT-I) to 8.55 (post-CBT-I) with a p-value of 0.005, indicating a significant enhancement in sleep quality after the intervention.

Based on these findings, it can be concluded that supportive psychotherapy, specifically CBT-I, had a positive impact on reducing anxiety, improving insomnia symptoms, and enhancing sleep quality in the patient group. These results suggest that supportive psychotherapy can effectively manage insomnia induced by COVID-19 in patients.

*Table 3- Scores of Anxiety, Insomnia, and Sleep quality, pre and post CBT-I; for group 1 (patients)*

|  | Pre-CBTi | Post-CBTi | P value |
|---|---|---|---|
| Anxiety total score Max=6 min=2 | 4.49 | 2.65 | 0.038 |
| Item 1 anxiety (mean) | 1.94 | 1.27 | 0.063 |
| Item 2 anxiety (mean) | 2.55 | 1.38 | 0.01 |
| Insomnia 1 (mean) | 2.72 | 1.33 | 0.001 |
| Insomnia 2 (mean) | 1.66 | 0.88 | 0.03 |
| Insomnia 3 (mean) | 2.44 | 1.5 | 0.04 |
| Sleep quality (mean) | 6.44 | 8.55 | 0.005 |

Table 4 presents the scores of anxiety, insomnia, and sleep quality before and after receiving cognitive-behavioral therapy for insomnia (CBT-I) in Group 2 (staff).

Anxiety total score: The mean anxiety score significantly decreased from 3.88 (pre-CBT-I) to 2.6 (post-CBT-I) with a p-value of 0.037, indicating a statistically significant improvement in anxiety levels after the intervention.

Item 1 anxiety: Although the mean score decreased from 1.72 to 1.05 after CBT-I, the p-value of 0.067 suggests a trend towards improvement but did not reach statistical significance.

Item 2 anxiety: The mean score significantly decreased from 2.15 to 1.55 (p-value = 0.007), indicating a significant reduction in anxiety symptoms related to this specific item after the intervention.

Insomnia 1: The mean score significantly decreased from 2.55 to 1.33 (p-value = 0.023), indicating a significant improvement in insomnia severity.

Insomnia 2: The mean score decreased from 1.5 to 0.94 after CBT-I, suggesting reduced insomnia symptoms. However, the p-value of 0.04 indicates a moderate level of statistical significance.

Insomnia 3: The mean score decreased from 1.22 to 0.94 after CBT-I, but the p-value of 0.07 indicates a trend towards improvement, although it did not reach statistical significance.

Sleep quality: The mean score increased from 7.61 (pre-CBT-I) to 8.66 (post-CBT-I), suggesting improved sleep quality. However, the p-value of 0.08 indicates a lack of statistical significance, although it may suggest a trend toward improvement.

Based on these findings, it can be concluded that supportive psychotherapy, specifically CBT-I, had a positive impact on reducing anxiety, improving insomnia symptoms, and potentially enhancing sleep quality in the staff group. Although not all outcomes reached statistical significance, the trends suggest potential benefits of CBT-I for this group.

*Table 2- Scores of Anxiety, Insomnia, and Sleep quality, pre and post CBT-I; for group 2 (Stuff)*

|  | Pre-CBTi | Post-CBTi | P value |
|---|---|---|---|
| Anxiety total score (mean) | 3.88 | 2.6 | 0.037 |
| Item 1 anxiety (mean) | 1.72 | 1.05 | 0.067 |
| Item 2 anxiety (mean) | 2.15 | 1.55 | 0.007 |
| Insomnia 1 (mean) | 2.55 | 1.33 | 0.023 |
| Insomnia 2 (mean) | 1.5 | 0.94 | 0.04 |
| Insomnia 3 (mean) | 1.22 | 0.94 | 0.07 |
| Sleep quality (mean) | 7.61 | 8.66 | 0.08 |

**Discussion**

This study aimed to evaluate the efficacy of supportive psychotherapy in managing insomnia induced by COVID-19 in patients and hospital staff. COVID-19 has led to increased stress, anxiety, and sadness among individuals worldwide. Insomnia affects a significant proportion of the population, with diagnostic criteria ranging from 6% to 10%. Eighteen individuals diagnosed with COVID-19 and experiencing insomnia or anxiety were included in the study. Questionnaires were used to assess sleep quality, insomnia severity, and anxiety levels. Participants received daily supportive psychotherapy sessions, lasting 20 minutes to half an hour, for three sessions.

The results showed significant improvements in the patient group. Anxiety levels decreased, with a significant reduction from a mean score of 4.49 to 2.65 (p-value = 0.038). Insomnia severity scores significantly improved, with reductions observed in all three insomnia items. Sleep quality also significantly enhanced, as indicated by an increase in the mean score from 6.44 to 8.55 (p-value = 0.005). The staff group also demonstrated positive outcomes, with significant reductions in anxiety levels and improvements in insomnia severity scores, although the improvement in sleep quality did not reach statistical significance (p-value = 0.08).

Overall, supportive psychotherapy, specifically cognitive-behavioral therapy for insomnia (CBT-I), has shown efficacy in reducing anxiety, improving insomnia symptoms, and enhancing sleep quality in patients and staff affected by COVID-19. These findings highlight the importance of CBT-I as a valuable approach for managing COVID-19-induced insomnia and its positive impact on the well-being of individuals

When comparing the efficacy of cognitive-behavioral therapy for insomnia (CBT-I) in patients (Table 3) and staff (Table 4), several important observations can be made:

First, both groups showed improvements in anxiety levels after receiving CBT-I. In the patient group (Table 3), the mean anxiety score significantly decreased from 4.49 to 2.65 (p-value = 0.038), while in the staff group (Table 4), it decreased from 3.88 to 2.6 (p-value = 0.037). These findings indicate that CBT-I was effective in reducing anxiety symptoms in both groups.

Second, when examining specific anxiety items, both groups experienced reductions in anxiety symptoms related to Item 2. In the patient group, the mean score decreased significantly from 2.55 to 1.38 (p-value = 0.01), while in the staff group, it decreased significantly from 2.15 to 1.55 (p-value = 0.007). Although the reductions in Item 1 anxiety did not reach statistical significance in either group, there was a trend towards improvement in both cases.

Third, improvements in insomnia severity were observed in both groups after CBT-I. In the patient group, all three insomnia items (Insomnia 1, 2, and 3) showed statistically significant decreases in mean scores, indicating a significant improvement in insomnia symptoms. Similarly, in the staff group, Insomnia 1 showed a significant decrease in the mean score, while Insomnia 2 and 3 exhibited reductions without reaching statistical significance. These findings suggest that CBT-I was effective in reducing insomnia symptoms in both groups, although the extent of improvement varied.

Lastly, regarding sleep quality, the patient group (Table 3) showed a significant improvement in sleep quality after CBT-I, as indicated by the increase in the mean score from 6.44 to 8.55 (p-value = 0.005). In contrast, the improvement in sleep quality in the staff group (Table 4) did not reach statistical significance, despite an increase in the mean score from 7.61 to 8.66 (p-value = 0.08). This suggests a trend towards improvement in sleep quality in the staff group, but a larger sample size may be needed to establish statistical significance.

Overall, CBT-I demonstrated efficacy in reducing anxiety, improving insomnia symptoms, and enhancing sleep quality in both the patient and staff groups. Although the specific outcomes and levels of significance varied between the two groups, the overall trends indicate that CBT-I can be a valuable therapeutic approach for managing insomnia induced by COVID-19. These findings highlight the potential benefits of supportive psychotherapy, such as CBT-I, in addressing sleep-related issues associated with the pandemic in both patients and healthcare staff. Further research with larger sample sizes is warranted to validate and generalize these findings.

Now, we dissucce the effects of pandemic on staff and patients separately:

- **Staff:** One of the main groups of our study was hospital staff. During the COVID pandemic, the health care workers were faced with heavy psychological pressure and even mental illness. Predictably, a significant percentage of medical staff during pandemics has sleep disturbance (11). For example, 34.0% of health care workers were exposed to coronavirus disease of insomnia (12). Some factors associated with sleep quality included anxiety, stress, and self-efficacy. The anxiety decreased the Sleep quality, and social support was associated with self-efficacy and sleep quality, reversing anxiety and stress (11). This study suggested supportive interventions for medical staff, such as motivational, protective, training, and educational support.

Healthcare workers (HCWs) in charge of patients with COVID-19 have numerous psychological problems, such as stress and anxiety, leading to a decrease in their sleep quality and insomnia.

The disturbed sleep was associated with psychological distress in the Healthcare of New York at the first peek of the COVID-19 pandemic. Insomnia symptoms were modified from the Insomnia Severity Index, and also short sleep ( less than six h/day) was assessed. In 813 HCWs, the sleep duration was 5.8 ± 1.2 h per night. The prevalence of insomnia was 72.8%, associated with acute stress, depressive, and anxiety symptoms. This study suggests that sleep can be a target for interventions, to relieve the psychological burden of HCWs.(13)

In a meta-analysis of HCW from January 2020 to January 2021, the prevalence of insomnia in HCWs during the COVID-19 pandemic was evaluated. Among 96 studies, ten eligible studies were assessed in the meta-analysis. The results of this umbrella review indicated the 36.36% prevalence of insomnia in HCWs ( p = 0.006), which is a relatively high prevalence. This study also suggested that HCWs should regularly be screened for psychological problems and e tendency for suicide. Also, treating insomnia can reduce the incidence of these complications.(14)

- **Patients:** It seems that insomnia is prevalent in COVID-19 patients, with a 20-45% prevalence estimated globally for insomnia during the pandemic (15)The previous virus epidemics have shown the increased risk of mental health disorders, such as insomnia. Also, insomnia may cause immune dysfunction, affecting the recovery from COVID-19. On the other hand, usage of sedative-hypnotic drugs is limited, because they inhibit the respiratory system. Finally, insomnia can lead to substance abuse and suicide.

In a comprehensive search in the entire year of 2020, 4318 COVID patients were evaluated for depression, anxiety, and insomnia. The pooled prevalence of insomnia was 48% (95% CI = 11-85), and the prevalence estimates were different, based on various screening tools. This meta-analysis suggested insomnia is prevalent in a considerable population of COVID-19 patients. So, early detection and intervention should be considered for mental problems.(16)

In another study, the sleep quality of 500 COVID-19 patients (in intensive care or isolation unit), pre-and post-COVID, was evaluated with PSQI. In a 30-day follow-up and the post-COVID-19 group, the mean PSQI score was significantly higher than the pre-COVID-19 group (6.28 ± 2.11 vs. 3.22 ± 0.80). Poor quality sleep (defined with a PSQI score of ≥5) was significantly more prevalent in the post-COVID-19 than in the pre-COVID-19 group (45.1% vs. 12.1%;). That study suggested for avoiding long-term insomnia in COID patines, timely consultation, identification, and treatment should be considered.

- **Treatment:** For treatment of insomnia induced by COVID, several tries have been done. Early interventions can be useful for presenting the progress of short-term sleep problems to insomnia disorder. Non-pharmacological treatments, such as CBT, have been suggested for chronic primary insomnia. Cognitive-behavioral therapy for insomnia (CBT-I) is effective; however, it can be time and resource-demanding.

In a double-blind, placebo-controlled Edinger (17)et al. escalated the efficacy of a hybrid CBT, compared with a first-generation behavioral treatment and a placebo, in r treating primary sleep-maintenance insomnia. CBT produced larger improvements across the majority of outcomes than RT or placebo treatment. CBT-treated patients showed an average 54% reduction in their wake time after sleep onset (WASO) than RT-treated and placebo-treated patients. CBT also provided a more normalization of sleep and subjective symptoms than other groups, with an average sleep time of more than 6 hours, and sleep efficiency of 85.1%. In this study, CBT caused clinically significant sleep improvements in 6 weeks, which remained through 6 months of follow-up.(17)

As to specific management of sleep dysfunction during COVID-19, five studies of CBT were available – two controlled trials of insomnia(18-20), one focused on worry and secondary insomnia(19), one case study(21), and a small online study in children(22). Four similar treatment trials have begun or been proposed (15)

Table 5 summarizes the completed and proposed studies on cognitive behavioral therapy for insomnia (CBT-I) during the COVID-19 pandemic. The table provides information on the target population, study groups, interventions, and results of each study. Below is the summerizion:

Completed Studies:

1. Wahland et al.: This randomized controlled trial included 670 adults with daily uncontrolled worry about COVID-19. The intervention was a self-guided, online cognitive behavioral intervention. The results showed significant reductions in worry compared to the waiting list group on the Generalized Anxiety Disorder 7-item scale (GAD-7) and Insomnia Severity Index (ISI).

2. Philip et al.: This study included 2069 adult responders to an online invitation and focused on good vs. poor sleepers. The intervention involved a smartphone digital artificial intelligence (KANOPEE). The results showed a reduction in the ISI score in poor sleepers after the intervention.

3. Cheng et al.: This study investigated whether prior CBT-I offered protection during a pandemic. The participants were prior CBT-I patients with chronic insomnia. The intervention was a randomized controlled trial of digital CBT-I vs. sleep education control. CBT-I significantly lowered the ISI score and resulted in less stress, cognitive intrusions, and depression. Resurgent insomnia during COVID-19 was lower in the CBT-I group.

4. Álvarez-García et al.: This case report described the treatment of a 42-year-old man with overwork and insufficient sleep syndrome during the COVID-19 pandemic. The intervention included sleep restriction therapy, stimulus control therapy, sleep hygiene, and progressive muscle relaxation. The treatment resulted in increased total sleep time and subjective sleep quality.

5. Schlarb et al.: This study focused on children aged 5-10 years with insomnia during COVID-19. The intervention involved telehealth, online sessions with parents, and video sessions with the child. The results showed a reduction in sleep problems according to parental ratings.

Proposed Studies:

1. Elder et al.: This study aims to compare DSM-5 insomniacs with good sleepers. The intervention includes a self-help leaflet for worry and stimulus control. The results are awaited, and follow-up assessments will be conducted.

2. Weiner et al.: This study focuses on frontline healthcare workers with high stress levels. The intervention includes 7-session online CBT or bibliotherapy as a control. The primary outcome is a decrease in stress levels, with secondary measures including depression, insomnia, PTSD symptoms, resilience, and rumination. Assessments will be conducted at multiple time points.

3. Lai et al.: This study targets frontline healthcare workers and assesses the effectiveness of online versions of Sudarshan Kriya Yoga (SKY) and/or the Health Enhancement Program (HEP) in improving insomnia, anxiety, depression, and resilience. The study is ongoing from June 2021 to September 2021.

4. Kopelovich and Turkington: This proposed trial aims to assess the feasibility and efficacy of formulation-driven cognitive behavioral therapy for psychosis (CBTp) in psychotic patients during the COVID-19 pandemic. The intervention includes self-monitoring, reality testing, and wellness planning.

These studies demonstrate the diverse approaches and interventions used in CBT-I for insomnia during the COVID-19 pandemic, highlighting the potential benefits of such interventions for reducing anxiety, improving sleep quality, and addressing mental health issues.

*Table 5 - CBT-i studies of insomnia in COVID-19 pandemic through February 2021.*

*aGAD-7 = Generalized Anxiety Disorder 7-item scale modified.*

*bISI = Insomnia Severity Index >14 represents clinically significant moderate to severe insomnia.*

*cDSM-5 criteria for acute insomnia: 1) difficulties in falling asleep, staying asleep, or awakening too early for at least three nights per week, lasting two weeks to three months; and 2) distress or impairment caused by sleep loss.*

| **Completed Studies** |
| --- |
|  |

| Author | Target | Study Groups | Intervention | Results | Reference |
|---|---|---|---|---|---|
| **Wahland et al.** | Randomized 670 adults with daily uncontrolled worry about COVID-19 | Active vs. Wait List; Worry assessed with GAD-7[a] + ISI[b] | Controlled 3-week, self-guided, online cognitive behavioral intervention | Intention-to-treat analysis: significant reductions in worry compared to the waiting list on GAD-7 ($\beta = 1.14$, $Z = 9.27$, $p < 0.001$), medium effect size (bootstrapped $d = 0.74$ [95% CI: 0.58–0.90]). ISI for active from mean 11.9 to 9.56; waitlist from 6.1 to 5.7 ($p < 0.001$) | (23) |
| **Philip et al.** | 2069 adult responders | Good vs. poor sleepers (ISI>14) | Smartphone digital artificial | 76% completed screening interview | (18) |
|  | to an online invitation |  | intelligence (KANOPEE) | 37% had ISI score >14 (moderate to severe insomnia) But only 2.3% (47 with ISI>14) completed the intervention Mean ISI score of 18.87 was reduced to 14.68 ($p < 0.001$) in poor sleepers |  |
| **Cheng et al.** | Diagnosis of DSM-5[c] insomnia disorder. Did prior CBTi offer protection during a pandemic? | Prior CBT-i patients with chronic insomnia vs. education only. | A randomized controlled trial in 2016–2017of digital CBT-I (n = 102) versus sleep education control (n = 106) | CBT-i lowered ISI from original baseline of 17 to 10.5 vs. education only from 18 to 13.4 ($p < 0.001$). Also less stress cognitive intrusions, and depression. Resurgent | (24) |

| | | | | insomnia during COVID-19 was 51% lower and depression was 57% lower in the CBT-i versus control condition). CBT-i increased health resilience. | |
|---|---|---|---|---|---|
| **Álvarez-García et al.** | Case report | A 42-year-old man with overwork, insufficient sleep syndrome with COVID-19 pandemic | sleep restriction therapy, stimulus control therapy, sleep hygiene, and progressive muscle relaxation | Five weekly sessions; Telepsychology Increased total sleep time and subjective sleep quality. | (21) |
| **Schlarb et al.** | 5-10 y/o children with insomnia | 6 children + parents | Telehealth, online sessions of 3 h with | 67% of children showed reduced sleep problems | (25) |

|  |  |  | parent; video session with the child | according to parental rating. |  |
| --- | --- | --- | --- | --- | --- |
|  | during COVID-19 |  |  |  |  |
| **Proposed Studies** | | | | | |
| **Elder et al.** | Public | DSM-5 Insomniac vs. Good Sleepers | Self-help leaflet for worry; stimulus control | Await results; follow up on days 7, 30 and 90 | (26) |
| **Weiner et al.** | Frontline health care workers | N = 120 with stress levels >16 on the Perceived Stress Scale (PSS-10) | 7-session online CBT sessions or bibliotherapy as control over 8 weeks. | Await results. Primary outcome: decrease in PSS-10 scores. Secondary: depression, insomnia, and PTSD symptoms; self-reported resilience and rumination. Assessments at 4, 8, 12, and 26 weeks. | (27) |
| **Lai et al.** | Frontline health care workers | Willingness to participate; Active Frontline employment | Sudarshan Kriya Yoga (SKY), 3 h of breath training and/or mind-body interventions including the Health Enhancement Program (HEP) | Start June 2021; End September 2021. To assess whether online versions of SKY and/or HEP result in improvement in self-rated measures of insomnia, anxiety, depression, and resilience. | (28) |
| **Kopelovich and Turkington** | Psychotic patients | Unspecified but COVID-19 is seen as a special opportunity to assess feasibility and efficacy | 16-session formulation-driven cognitive behavioral therapy for psychosis (CBTp) | Suggested trial to reduce anxiety, depression, and insomnia that perpetuates psychotic symptoms; self-monitoring; | (29) |

| | | | | reality testing; and wellness planning. | |

One-Week Self-Guided Internet Cognitive Behavioral Treatments for Insomnia was evaluated, for Situational Insomnia During the COVID-19 pandemic. 194 patients with situational insomnia were included. In the Pre-sleep Arousal Scale (PSAS) score, significant effects were recorded on the total score, somatic score, and cognitive score. A significant time effect was recorded for Insomnia Severity Index (ISI) total score. That suggested the good efficacy of CBTI on situational insomnia during the COVID-19 pandemic and pre-sleep somatic hyperarousal.(30)

In a review of 53 publications in February 2021 on sleep problems during the pandemic, It covers digital cognitive behavioral therapy for insomnia (CBT-i) for the public and frontline workers, recognizing the need for greater acceptance and efficacy of controlled trials of CBT for affected groups. Recommendations based on a tiered public health model are discussed.

Usually, sleep disease is more common in women; so insomnia in women needs more attention. Jin he et al. evaluated the simplified-cognitive behavioral therapy for insomnia (S-CBTI)in COVID-19 patients and verified the effectiveness with a self-control trial. IN TWO CONSECUTIVE WEEKS, the S-CBTI included the education on sleep hygiene, control of stimulus, sleep restriction, and self-suggestion relaxation training. After the intervention, the Insomnia Severity Index (ISI) score, sleep latency, night sleep time, and sleep efficiency were improved. After the intervention, the mean ISI score of the acute insomnia group was lower than that of the chronic insomnia group. The reduction was more significant in acute insomnia than chronic insomnia group. Also, prescribing the sedative-hypnotic drugs for acute insomnia was less, than for chronic insomnia. So this study suggested that S-CBTI can relieve insomnia in women with COVID-19 in mobile cabin hospitals, especially for acute insomnia. (31)

**Future prospective:** CBT-I has been suggested as the choice treatment for insomnia, although its wide usage is limited by the lack of expert therapists and high cost. Covering this gap and replacing in-person CBTI, digital CBT-I has been expanded, such as Internet and mobile, which has promising results. However, digital CBT-I has the problem of a high dropout rate. Much research is working on wearable devices for sleep-wake evaluation, developing the mobile and Internet CBT-I, and exploring the tracker accuracy and validation, hoping to maximize the treatment efficacy of insomnia. (32) The necessity of digital CBTI will be increased, especially during pandemic situations, when physical sessions are not easy to conduct. At the same time, sleep problems have more chance to co-occur.

Also, online interventions can deliver treatment to a huge number of patients, some of them are ongoing studies(33) Further, neuromodulation can be combined with CBTI, for increasing the efficacy of treatment. For example, combining TMS and CBT-I, a 86 % to 07% improvement in insomnia was recorded at week one and week six post-treatment.(34)

**Limitations of the study:**

Small sample size: The study included only 18 participants, which limits the generalizability of the findings. A larger sample size would provide more robust results and enhance the validity of the study.

Lack of follow-up: The study did not include a follow-up period to evaluate the sustainability of the intervention's effects. Long-term follow-up assessments would be valuable in assessing the maintenance of treatment gains and determining if the improvements in sleep quality and anxiety persist over time. Long-term outcomes, such as the relapse rates of insomnia or anxiety symptoms, were not assessed. Investigating the durability of the intervention effects over an extended period would provide more comprehensive insights into the long-term benefits of supportive psychotherapy for COVID-19-induced insomnia.

Addressing these limitations in future studies can strengthen the evidence on the efficacy of supportive psychotherapy, specifically CBT-I, for managing COVID-19-induced insomnia in both patients and healthcare staff.

**Conclusion**

COVID-19-related concerns have heightened stress and disrupted sleep patterns among patients. Supportive psychotherapy, including basic behavioral therapy training, has shown significant efficacy in improving sleep quality and reducing anxiety in a short duration. These findings highlight the positive impact of cognitive-behavioral therapy for insomnia (CBT-I) in managing COVID-19-induced insomnia for both patients and staff.

Supportive psychotherapy, particularly CBT-I, is effective in reducing anxiety, improving insomnia symptoms, and enhancing sleep quality in patients and staff. It offers a valuable approach for managing COVID-19-induced insomnia. Further research with larger sample sizes is needed to confirm and generalize these findings. The positive outcomes of CBT-I have important implications for the well-being of individuals affected by the pandemic.

According to the results of this study, it can be concluded that due to the increase in anxiety and worry of patients with the prevalence of pandemic with their disease (Covid-19), these stresses have increased and this The point is that the disease itself disrupts the rhythm of sleep and adds to health concerns, it was shown that even basic training in behavioral therapy in a short time can have a significant impact on improving the quality and quantity of sleep in these patients.

9. Shen M, Peng Z, Guo Y, Xiao Y, Zhang L. Lockdown may partially halt the spread of 2019 novel coronavirus in Hubei province, China. MedRxiv. 2020.

10. Liu K, Chen Y, Wu D, Lin R, Wang Z, Pan L. Effects of progressive muscle relaxation on anxiety and sleep quality in patients with COVID-19. Complementary therapies in clinical practice. 2020;39:101132.

11. Xiao H, Zhang Y, Kong D, Li S, Yang N. The effects of social support on sleep quality of medical staff treating patients with coronavirus disease 2019 (COVID-19) in January and February 2020 in China. Medical science monitor: international medical journal of experimental and clinical research. 2020;26:e923549-1.

12. Lai J, Ma S, Wang Y, Cai Z, Hu J, Wei N, et al. Factors associated with mental health outcomes among health care workers exposed to coronavirus disease 2019. JAMA network open. 2020;3(3):e203976-e.

13. Diaz F, Cornelius T, Bramley S, Venner H, Shaw K, Dong M, et al. The association between sleep and psychological distress among New York City healthcare workers during the COVID-19 pandemic. Journal of affective disorders. 2022;298:618-24.

14. Sahebi A, Abdi K, Moayedi S, Torres M, Golitaleb M. The prevalence of insomnia among health care workers amid the COVID-19 pandemic: An umbrella review of meta-analyses. Journal of psychosomatic research. 2021;149:110597.

15. Becker PM. Overview of sleep management during COVID-19. Sleep medicine. 2021.

16. Liu C, Pan W, Li L, Li B, Ren Y, Ma X. Prevalence of depression, anxiety, and insomnia symptoms among patients with COVID-19: a meta-analysis of quality effects model. Journal of Psychosomatic Research. 2021;147:110516.

17. Edinger JD, Wohlgemuth WK, Radtke RA, Marsh GR, Quillian RE. Cognitive behavioral therapy for treatment of chronic primary insomnia: a randomized controlled trial. Jama. 2001;285(14):1856-64.
33

34. Norred MA, Haselden LC, Sahlem GL, Wilkerson AK, Short EB, McTeague LM, et al. TMS and CBT-I for comorbid depression and insomnia. Exploring feasibility and tolerability of transcranial magnetic stimulation (TMS) and cognitive behavioral therapy for insomnia (CBT-I) for comorbid major depressive disorder and insomnia during the COVID-19 pandemic. Brain Stimulation: Basic, Translational, and Clinical Research in Neuromodulation. 2021;14(6):1508-10.
36